\documentclass[journal]{IEEEtran}
\usepackage{color}
\usepackage{graphicx}
\usepackage{amsthm}
\usepackage{amsmath}
\usepackage{amssymb}
\usepackage{booktabs}
\usepackage{setspace}
\usepackage{textcomp}
\usepackage{multirow}
\usepackage{makecell}
\usepackage{balance}
\usepackage{xcolor}
\usepackage{float}
\usepackage{subfigure}
\usepackage{lineno}
\begin{document}
\title{A time-efficient frame size adjustment based approach for RFID anti-collision}

\author{Jian~Su,~\IEEEmembership{member,~IEEE},~Kexiong~Liu,~haipeng~Chen,~and~Yu~Han
\thanks{J. Su (corresponding author) is with Nanjing University of Information Science and Technology, Nanjing 210044, China (e-mail: sj890718@gmail.com).} 
\thanks{K. Liu is with Beijing Forestry University, Beijing 100083, (e-mail: kexiongliu@bjfu.edu.cn).}%
\thanks{H. Chen is with Northeast Electric Power University, Jilin 132012, (e-mail: haipeng0704@126.com).}%
\thanks{Y. Han is with University of Electronic Science and Technology, Chengdu 611731, (e-mail: hanyu911018@126.com).}%
\thanks{Digital Object Identifier xxxx}
}

\maketitle

\begin{abstract}
Fast and efficient identify a large number of RFID tags in the region of interest is a critical issue in various RFID applications. In this paper, a novel sub-frame-based algorithm with a time-efficient frame size adjustment strategy to reduce the time complexity for EPCglobal C1 Gen2 UHF RFID standard is proposed. By observing the slot statistics in a sub-frame, the tag quantity is estimated by the reader, which afterwards efficiently calculates an optimal frame size to fit the unread tags. Only when the expected time efficiency in the oncoming frame is higher than that in the previous frame, the reader starts the new identification round with the updated frame. Moreover, the estimation of the proposed algorithm is implemented by the look-up tables, which allows dramatically reduction in the computational complexity. Simulation results show noticeable throughput, time efficiency, and identification speed improvements of the proposed solution over the existing approaches.
\end{abstract}

\begin{IEEEkeywords}
RFID, anti-collision, sub-frame, time efficiency.
\end{IEEEkeywords}

\IEEEpeerreviewmaketitle

\section{\textsc{Introduction}}
\IEEEPARstart{R}{adio} frequency identification (RFID) is a non-contact information collection technology, which can automatically identify and read the object data through radio frequency signal. RFID enables fast non-line-of-sight, mobile, multi-object recognition, position and tracking. It has been widely used in intelligent management and monitoring of people, object and asset in various fields penetrating into our daily lives because of its long identification distance, fast speed, and the tag's large memory capacity and reusability, etc. In various RFID application systems, a reader usually need to quickly and accurately identify a mass of tags within its coverage. Since the reader communicates with the tags via a shared wireless channel, if more than a tag respond to the reader at the same time, information collision occurs [2]. Such phenomenon is termed as multi-tag collision, which leads to reduced identification efficiency, increased omission ratio and identification latency, and eventually limits the RFID applications. Therefore, to tackle such collision problems and minimize their impact, a high-efficiency anti-collision algorithm must be employed by RFID systems. The commonly used anti-collision protocols include Aloha-based and Tree-based ones.

In tree-based algorithms, colliding tags are iteratively split into subsets by using channel feedback, in the end, each subset contains a tag at most. Aloha-based protocols divide the time into several frames contain multiple time intervals (called slot), in order to respond to the reader tags pick one slot in every frame randomly. As a most widespread standard in RFID systems, EPCglobal C1 Gen2 [3] adopts a dynamic frame slotted Aloha (DFSA) protocol to manage the identification process of tags. Recently, most commercial RFID systems are based on EPCglobal C1 Gen2, raised more concerns on DFSA protocols [3-5].

Specifically, DFSA protocol is characterized by the strategy that employs to adapt the frame size along identification process. The performance of DFSA is determined by both the cardinality (the number of unread tags) estimation as well as setting of frame size. Most previous solutions require a vast computation costs so that the accuracy of estimation can be ensured. However, most RFID readers in practice are structured with a single-chip microprocessor, which has limited computational ability and memory. Recently, plenty of state-of-the-arts research have proposed energy-efficient DFSA algorithms in order to reduce computational overhead. The literature [6] presented an anti-collision protocol which depends on just one examination of current frame size at a specific time slot during each identification round. Solic et al. [7] introduced an Improved Linearized Combinatorial Model (ILCM) to estimate the cardinality at the cost of modest calculation. Since the ILCM adopts a frame-by-frame estimation based on the number of idle, success, and collision slots observed in the previous full frame, the performance of ILCM is limited to the accuracy of a single estimation. Therefore, its performance fluctuates sharply accompanied by the tag number varies by a big margin. To achieve the robust performance, the slot-by-slot version of ILCM has been presented in [8]. In [9], the authors presented a FuzzyQ method which integrates fuzzy logic with a DFSA algorithm. A fuzzy rule based system is defined to model the current frame size and the collision or idle response rate as fuzzy sets to adaptively calculate frame size. However, the performance of FuzzyQ is need to further improved.

Sub-frame based algorithms [10-11] were only recently proposed. In literature [10], the observation of ratio relations between idle and collision statistics of current sub-frame can be used to predict the tag quantity. Nevertheless, since the usage of empirical correlation not theoretical calculation, the accuracy of estimation is not enough. In [11], by observing current sub-frame, the MAP estimation function is used to estimate the tag quantity. The DS-MAP picks up an estimation result from the look-up tables instead of calculating a real-time result of the tag number during the identification process. Although the DS-MAP can decrease the total number of slots during the identification process, it doesn't consider the duration discrepancy of different slot type. Therefore, the DS-MAP algorithm is inefficient in terms of time efficiency or identification time.

In order to guarantee the computationally efficient enough and reduce the identification time during the identification process, we present an anti-collision solution namely time-efficient frame adjustment strategy (TEFAS) based algorithm. The proposed algorithm integrates the low-cost estimation method, adaptive frame size calculation strategy and efficient frame size adjustment policy. To be specific, the presented algorithm ascertains the optimal frame size based on both estimated tag quantity and the time duration of a slot. Moreover, the proposed frame size adjustment policy can avoid the improper frame adjustment. The results of extensive simulation indicate that the presented algorithm can perform better than the reference methods.

\IEEEpeerreviewmaketitle

\section{\textsc{Algorithm Description}}
\subsection{\textsc{Tag Quantity Estimation Strategy}}
In most RFID application scenarios, number of tags remains unknown to the reader in advance, hence the reader need to estimate the tag number accurately to maximize the performance of the proposed algorithm. Here we also refer to the maximum a posteriori probability (MAP) method to calculate the cardinality of tag population based on feedback from a sub-frame. Although MAP can achieve an accurate estimation, its high computational overhead impedes its application in low-cost RFID platforms. In the proposed estimation method, we design look-up tables to pre-store the estimation results. Restricted by the sub-frame size and the item quantity in the tables, the proposed estimation strategy is space-efficient and implementable. Considering $n$ tags allocated in $F$ slots, the probability that idle slot occurs $e$ times, success slot occurs $s$ times, and collision slots occurs $c$ times in a sub-frame $F_{sub}=e+s+c)$ (where $F_{sub}$ is a proportion of a full frame) can be expressed as
\begin{equation}
P\left( {n\left| {e,\;s,\;c} \right.} \right) = \frac{{F_{sub} !}}{{e!s!c!}}P_i^e P_s^s P_c^c
\end{equation}
where $P_i$, $P_s$, and $P_c$ are the probabilities of idle, success and collision slot occurs in the full frame, respectively. The tag quantity involved in a sub-frame is determined when the value of $P\left( {n\left| {e,\;s,\;c} \right.} \right)$ is maximized. So, the estimation result in a sub-frame is $\hat n_{sub}$. Then the estimated tag quantity involved in the full frame can be calculated as
\begin{equation}
\hat n_{est}  = \hat n_{sub}  \times \frac{F}{{F_{sub} }}
\end{equation}

To reduce computational complexity, the estimated results of the tag cardinality during the sub-frame can be stored in the preset tables. The recommendation setting of $F_{sub}$ can be referred to [10-11] and can be listed in Tab. I.
\begin{table}[htbp] \centering
\caption{\textsc{The Recommendation Setting of $F_{sub}$}}
\begin{tabular}{cccccc}
\toprule
$F$ & 8$\sim$16 & 32$\sim$64 & 128$\sim$256 & 512$\sim$1024 & \textgreater 1024 \\
\midrule
$F_{sub}$ & 4 & 8 & 16 & 32 & 64 \\
\bottomrule
\end{tabular}
\end{table}

\subsection{\textsc{Adaptive Frame Size Calculation}}
The most existing DFSA algorithms set the frame size as the nearest value to the estimated tag number. Unlike the conventional DFSA algorithms, the proposed algorithm calculates the frame size by maximize the time efficiency $T_{effi}$ to identify all tags, which can be defined as [6][10]
\begin{equation}
T_{effi}  = \frac{{S \cdot T_S }}{{T_S  \cdot S + T_E  \cdot E + T_C  \cdot C}}
\end{equation}
where $E$, $S$, and $C$ are the statistics of idle slots, success slots and collision slots during the whole identification process, respectively. $T_E$, $T_S$, and $T_C$ are the time intervals of above three slot types, and can be expressed in [10].

Considering the number of tags is $n$ and the initial frame size is $F$, the fill level of $r$ tags allocated in a slot is described by a binomial distribution with $1/F$ occupied probability:
\begin{equation}
P_r  = C_n^r \left( {\frac{1}{F}} \right)^r \left( {1 - \frac{1}{F}} \right)^{n - r}
\end{equation}

Accordingly, $P_i=P_{r = 0} $, $P_s=P_{r = 1}$, and $P_c=P_{r \textgreater 1}$ are the corresponding probabilities that a slot is idle, successful and collided, respectively. If $F$ is assumed large enough, the distribution of tags can be approximated as Poisson distribution with mean $\lambda=n/F$. Then, the parameters $E$, $S$, and $C$ in Eq. (3) can be approximated as
\begin{equation}
E = F \cdot P_i  = F\left( {1 - \frac{1}{F}} \right)^n  \approx F \cdot e^{ - \lambda }
\end{equation}
\begin{equation}
\begin{array}{l}
 S = F \cdot P_s  = F \cdot \frac{n}{F}\left( {1 - \frac{1}{F}} \right)^{n - 1}  \\
 \;\;\; \approx F \cdot \lambda  \cdot \left( {\frac{F}{{F - 1}}} \right)e^{ - \lambda }  \\
 \end{array}
\end{equation}
\begin{equation}
C = F \cdot P_c  = F \cdot \left( {1 - P_i  - P_s } \right)
\end{equation}

Substitute Eqs. (5)-(7) into Eq. (3), the $T_{effi}$ can be approximated as
\begin{equation}
T_{effi}  \approx \frac{{\lambda e^{ - \lambda }  \cdot T_S }}{{\lambda e^{ - \lambda }  \cdot T_S  + T_E  \cdot e^{ - \lambda }  + T_C  \cdot \left( {1 - (1 + \lambda )e^{ - \lambda } } \right)}}
\end{equation}

We take the first derivative of $T_{effi}$ with respect to $\lambda$. Let the derivative equals to zero, we then have
\begin{equation}
\frac{{dT_{effi} }}{{d\lambda }} = \frac{{T_S  \cdot \left( {T_E  - T_C \left( {e^\lambda  \left( {\lambda  - 1} \right) + 1} \right)} \right)}}{{\left( {T_E  - T_C  + T_S  \cdot \lambda  - T_C  \cdot \lambda  + T_C  \cdot e^\lambda  } \right)^2 }} = 0
\end{equation}

The simple bisection search or Newton's methods can be used to solve the above non-linear equation of one variable, and transforming the Eq. (9), we can have
\begin{equation}
e^\lambda  \left( {\lambda  - 1} \right) + 1 = \frac{{T_E }}{{T_C }}
\end{equation}

By solving the Eq. (10), the value of $\lambda$ to maximize the $T_{effi}$ can be calculated as
\begin{equation}
\lambda  = 1 + W\left( {(\frac{{T_E }}{{T_C }} - 1)e^{ - 1} } \right)
\end{equation}
where $W(*)$ denotes as the Lambert W-function. Since $\frac{{d^2 T_{effi} }}{{d\lambda ^2 }} < 0$, consequently, the optimal frame size in the proposed algorithm can be set as
\begin{equation}
F_{opt}  = \frac{{\hat n_{est} }}{\lambda }
\end{equation}

\subsection{\textsc{Frame Size Adjustment Strategy}}
The mainstream frame size adjustment strategy can be divided into three categories. First is Frame-by-Frame (FbF) in which the reader calculates the new frame size at the end of the current frame. The FbF strategy is not efficient when the frame size is far away from the number of tags. Second is Slot-by-Slot (SbS) in which the reader calculates the new frame size at every slot of the current frame. The SbS strategy suffers from a rather high complexity. Finally, the sub-frame solution provides the flexibility of ending the current frame in advance to maintain the performance stability with a reduced computational complexity. In our proposed algorithm, we adopt a hybrid strategy combining sub-frame observation and SbS. At every slot, the reader keeps track of the relation between $E$ and $C$. And then the reader will reset the sub-frame size if the difference value between $E$ and $C$ is above the threshold value. After the reading of $F_{sub}$ slots, the reader estimates the tag quantity according to Eq. (2). And then the new frame size for the next identification round can be given by Eq. (15). Then the reader computes the $T_{effi1}$ and $T_{effi2}$ with the current frame size and the new frame size, respectively. The policy is to end the current frame and to adopt the new frame size if $T_{effi1} \textless T_{effi2}$. Otherwise, the reader will go to the next slot of the current frame. The identification process ends until no collision occurs. According to the hybrid frame size adjustment strategy, the algorithm can achieve a better and stable performance.

\subsection{\textsc{The Proposed Tag Identification Algorithm}}

Combining the tag quantity estimation, frame size calculation, and frame size adjustment strategy, we propose the anti-collision algorithm TEFAS as follows.
\noindent\rule{80mm}{1pt}
\begin{spacing}{0.8}
\textbf{Algorithm TEFAS} \emph{Reader Operation}\\
\noindent\rule[0.25\baselineskip]{80mm}{0.5pt}\\
1: Initialize $F_{curr}=F_{ini}$, $F_{sub}$, $E$, $S$, and $C$ \\
2: \textbf{while} (unidentified tags ¡Ù0) \\
3: \qquad The reader identifies tags among $F_{curr}$ \\
slots and counts $E$, $S$, and $C$ slot by slot. \\
4: \qquad\textbf{if} $E - 3.2*C/\lambda \textgreater threshold$ \\
5: \qquad\quad $F_{sub}=i$; \\
6: \qquad\textbf{else if}$E - 0.6*C/\lambda \textless -threshold$ \\
7: \qquad\quad $F_{sub}=i$; \\
8: \qquad\textbf{end if}\\
9: \qquad\textbf{if} $i==F_{sub}$ \\
10: \quad\quad Estimate the tag quantity and calculate the new \\
frame size by using Eqs. (2) and (12). \\
11: \quad\quad\textbf{if} $T_{effi1} \textless T_{effi2}$ \\
12: \quad\quad\quad Update the $F_{curr}$ according to Eq. (12) and \\
update $F_{sub}$ according to Tab. I. \\
13: \quad\quad\textbf{else} \\
14: \quad\quad\quad i++; \\
15: \quad\quad\textbf{end if}  \\
16: \quad\textbf{else} \\
17: \quad\quad i++; \\
18: \quad\textbf{end if} \\
19: \textbf{end while} \\
\normalsize
\noindent\rule{80mm}{1pt}
\end{spacing}

\noindent where $threshold=Muliply*Q$ is an upper value that allows the current frame to be end in advance. If the relative number of $E$ vs. the adjusted number of $C$ falls within the $threshold$, the frame size is unchanged. Otherwise, the reader ends the ongoing frame. In our simulations, the $Multiply$ is set as 4.

\section{\textsc{Simulation Results}}
The identification performance of the proposed algorithm and the reference methods including FuzzyQ [9], SUBF-DFSA [10], ILCM [7], and Impinj R2000 [5] was compared by carrying out extensive Monte Carlo simulations. Various metrics including system throughput, time efficiency, average identification time to identify one tag are taken into account to evaluate the performance of TEFAS. The primary time parameters refer to the literatures [10-11]. All simulations are carried out by using MATLAB 2010. Each simulation has been run 1000 times and averaged the result.
\begin{figure}[!tbp]
\centering
\includegraphics[width=8.57cm, height=6.4cm]{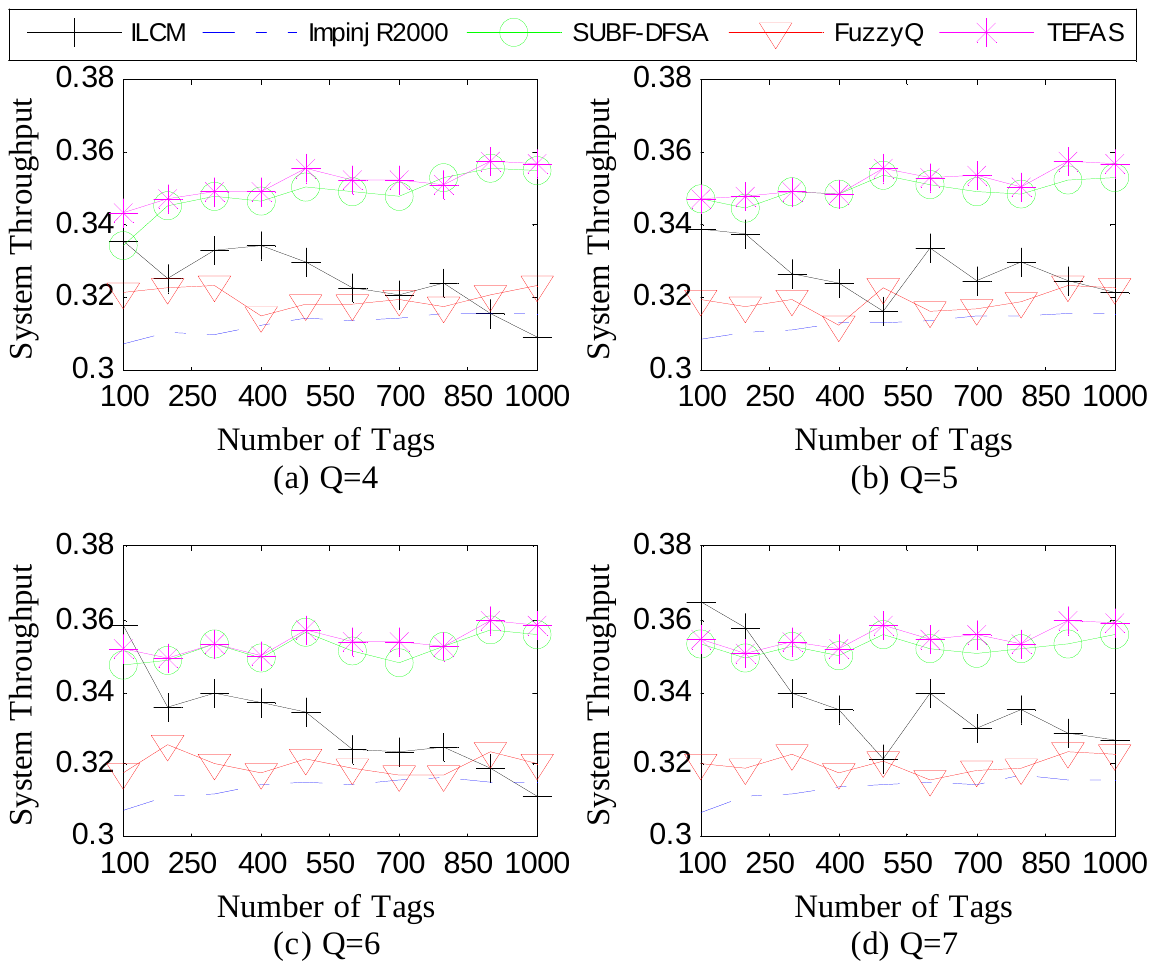}
\caption{Comparison of system throughput for various algorithms}
\label{Fig.1}
\end{figure}

Fig. 1 compares the system throughput of various algorithms under different initial frame size. As can be found from Fig. 1, the performance of ILCM is most sensitive to the initial frame size than other algorithms. When the number of tags is much larger than the size of frame, the ILCM can hardly adjust to a proper frame size to fit the unread tags, which will cause the degradation of performance. That is to say, the stability and scalability of ILCM is difficult to get used to a widely varying of the tag population. Since the other algorithms adopt the in-frame-like frame adjustment mechanism that allows the frame to be end in advance, so they can guarantee more stable performance. Also can be seen from Fig. 1, the average system throughput of five algorithms from the highest to the lowest are TEFAS, SUBF-DFSA, ILCM, FuzzyQ, and Impinj R2000. Although the Impinj R2000 and FuzzyQ can improve the performance stability and reduce the computational complexity, their system throughput are below that of ILCM. Specifically, the average system throughput of TEFAS is 0.3533, achieves a 6.9\% improvement over ILCM.
\begin{figure}[!tbp]
\centering
\includegraphics[width=8.57cm, height=6.4cm]{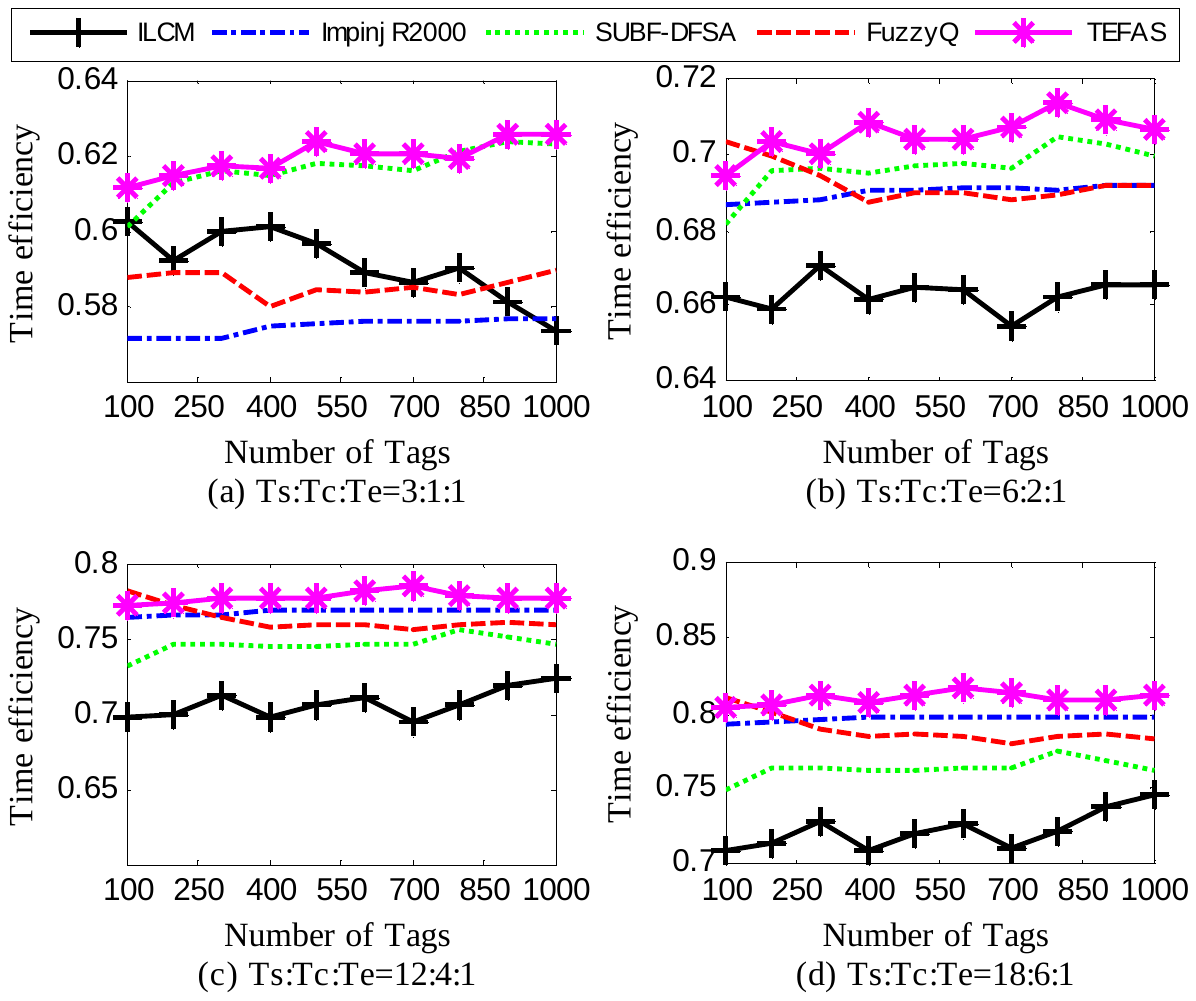}
\caption{Comparison of time efficiency for various algorithms}
\label{Fig.2}
\end{figure}

\begin{figure}[!tbp]
\centering
\includegraphics[width=8.57cm, height=6.4cm]{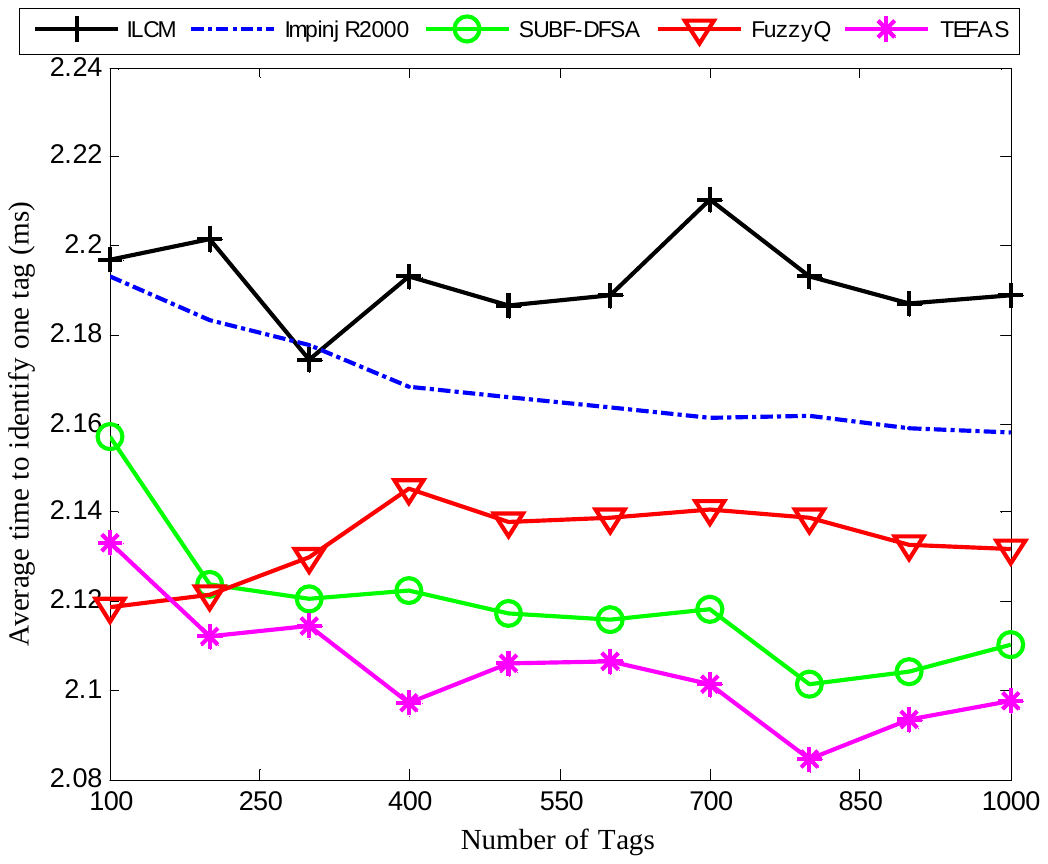}
\caption{Comparison of average identification time for various algorithms}
\label{Fig.3}
\end{figure}

In order to illustrate the advantage of the TEFAS, we show the time efficiency of various algorithms in Fig. 2. The metric time efficiency is dependent on the timing parameters of EPCglobal C1 Gen2, particularly on the intervals of the slots, $T_E$, $T_S$, and $T_C$, respectively. For this purpose, the reference methods are evaluated for different ratio between $T_C$ and $T_E$. All algorithms present an increasing time efficiency with the increasing between $T_C$ and $T_E$. As can be seen various algorithms show the discrepant time efficiency under the different conditions. For example, the time efficiency of Impinj R2000 is the lowest when $T_C=T_E$. As the increase in $T_C/T_E$, the Impinj R2000 achieves a significant performance improvement. Since the frame size can be adaptively adjusted to adapt to the different $T_C/T_E$, the TEFAS can always hold the best time efficiency compared to other algorithms.

Fig. 3 illustrate the average simulation identification time so as to distinguish one tag with an initial frame size of 16. Note that the average identification time for identifying one tag includes the coordination time used to transmitting command and guard time beside the necessary time used for message (such as ID or EPC) transmission. The average time is calculated by Eqs. (3)-(6). As can be observed from Fig. 3, the TAFSA consumes the least average time to identify one tag. It spends about 2.1045 ms, i.e., an identification speed is of 475 tags/s. This means that the TEFAS could identify more tags within unit time. Also, since our presented solution is on the basis of the same hardware platform of EPCglobal C1 Gen2 standard, there won't be introduced the extra cost.

\section{\textsc{Conclusion}}
In this paper, we presented a time-efficient anti-collision protocol to identify multiple RFID tags within one reader's filed. The proposed scheme combine the sub-frame observation and slot-by-slot to optimize the current frame size. According to the adaptive adjustment mechanism, the TEFAS can restrain the performance degradation in time efficiency due to the duration discrepancy. Moreover, the tag quantity estimation is implemented by the look-up tables, which allows dramatically reduction in the computational complexity. The results of simulation illustrate that the presented TAFSA outperforms other algorithm in various metrics.

\ifCLASSOPTIONcaptionsoff
  \newpage
\fi

\end{document}